\documentclass[pre,twocolumn,reprint,showpacs]{revtex4}
\usepackage{amsmath,amssymb,braket}
\usepackage{graphicx}

\newcommand{\dd}{\mathrm{d}}
\newcommand{\ii}{\mathrm{i}}
\newcommand{\ee}{\mathrm{e}}

\begin{document}
\title{
Finite-Size Scaling Analysis of
the Eigenstate Thermalization Hypothesis
in a One-Dimensional Interacting Bose gas
}
\date{\today}
\author{Tatsuhiko N. Ikeda$^1$, Yu Watanabe$^2$, and Masahito Ueda$^1$}
\affiliation{$^1$Department of Physics, University of Tokyo, 7-3-1 Hongo, Bunkyo-ku,
Tokyo 113-0033, Japan\\
$^2$Yukawa Institute for Theoretical Physics, Kyoto University,
Kitashirakawa Oiwake-Cho, 606-8502 Kyoto, Japan}

\begin{abstract}
By calculating correlation functions for the Lieb-Liniger model based on the
algebraic Bethe ansatz method,
we conduct a finite-size scaling analysis of the eigenstate thermalization hypothesis (ETH)
which is considered to be a possible mechanism of thermalization in isolated quantum systems.
We find that the ETH in the weak sense holds in the thermodynamic limit even for an integrable system
although it does not hold in the strong sense.
Based on the result of the finite-size scaling analysis,
we compare the contribution of the weak ETH to thermalization
with that of yet another thermalization mechanism, the typicality, 
and show that the former gives only a logarithmic correction to the latter.
\end{abstract}


\maketitle
{\it Introduction.---}
Recently,
there has been a resurgence of interest in 
understanding thermalization
from quantum mechanics~\cite{Polkovnikov11,Yukalov12} due in part to
experimental techniques
in ultracold atomic gases
which enable one to prepare,
control, and measure
isolated quantum systems~\cite{Kinoshita06,Trotzky11},
and also due to
several theoretical advances~\cite{
Tasaki98,Goldstein06,
Reimann08,Rigol08,Gogolin11,Ikeda11}.
Among possible mechanisms for thermalization,
the eigenstate thermalization hypothesis (ETH)
has attracted much attention~\cite{Deutsch91,Srednicki94,Rigol08}.
The ETH states that
the expectation value of an observable stays the same over
manybody eigenstates having close eigenenergies
in the thermodynamic limit.
The validity of the ETH has been examined
by using the numerical diagonalization
of the Hamiltonian~\cite{
Horoi95,Rigol08,Rigol09,Rigol10,Santos10}.
It has been claimed that
the ETH holds for the case in which
the system is non-integrable or chaotic~\cite{Rigol08,Rigol09,Rigol10,Santos10}.
These studies focus on the dependence
on the characteristics of the system
such as geometrical configurations
and parameter sets,
whereas the size of the system cannot be changed sufficiently
due to an exponential growth of the numerical cost.
However, since the ETH concerns the thermodynamic limit,
it is essential to analyze the finite-size scaling of the ETH.

In this Rapid Communication,
we identify the finite-size scaling properties of the ETH
by calculating the correlation functions over each manybody eigenstate
for the Lieb-Liniger model~\cite{Lieb63}
in which the Bethe ansatz method allows us to obtain the exact manybody eigenstates
and thus to circumvent the numerical difficulty in diagonalizing the Hamiltonian.
We show
the ETH holds in the weak sense but does not in the strong sense.
(The definitions of weak and strong will be given below~\cite{Biroli10}.)
In particular,
the weak ETH becomes better satisfied for larger systems
according to the power law
in the number of particles of the system.
This suggests that there exist situations in which
the microcanonical ensemble is applicable
to integrable systems in the thermodynamic limit,
while the generalized microcanonical or Gibbs ensemble leads to
better predictions in such systems including the Lieb-Liniger model~\cite{
Rigol06,Rigol07,Cassidy11,Barthel08,Cazalilla06,Iucci09,Calabrese07,Eckstein08,Kollar08,Mossel12}.
Then,
by using the result of the finite-size scaling of the weak ETH,
we discuss the quantitative relation between the ETH
and the typicality~\cite{Popescu06,Goldstein06,Sugita06,Reimann07},
the latter being yet another possible mechanism.
of thermalization in isolated quantum systems.
We find that the ETH contributes to thermalization
at most as a logarithmic correction to the typicality.

{\it Formulation of the Problem.---}
We consider an isolated quantum manybody system whose Hamiltonian is $H$.
We denote each eigenstate of $H$ with eigenenergy $E_i$ as $\ket{E_i}$
and expand the initial state $\ket{\psi_0}$ in terms of them:
$\ket{\psi_0} = \sum_i C_i \ket{E_i}$.
Then, the state at time $t$ is given by $\ket{\psi_t}=\sum_i C_i \ee^{-\ii E_i t/\hbar}\ket{E_i}$.
Here, we assume the condition of non-degenerate energy gaps~\cite{Short11} which states that
if $E_i-E_j = E_k -E_l \neq 0$, then $E_i=E_k$ and $E_j=E_l$.
Then, according to the study of equilibration~\cite{Reimann08,Short11},
if the inverse participation ratio $Q^{-1} \equiv \sum_i |C_i|^4$ is sufficiently large,
the time-dependent expectation value of an observable $A$
relaxes to its long-time average:
\begin{equation}
\braket{\psi_t|A|\psi_t} \longrightarrow \langle A \rangle^\text{LT}
\equiv \lim_{T\rightarrow \infty} \int_0^T\frac{\dd t}{T} \braket{\psi_t|A|\psi_t}
\end{equation}
which means that the fluctuation of $\braket{\psi_t|A|\psi_t}$ around $\langle A \rangle^\text{LT}$
becomes negligible in the long run~\cite{Reimann08}.
We note that the long-time average of $A$ can be rewritten
as $\langle A \rangle^\text{LT}=\sum_i |C_i|^2 \braket{E_i|A|E_i}$
unless the Hamiltonian $H$ has degeneracy
(see Ref.~\cite{Short11} for the generalization to the case in which $H$ has degeneracies).
This is because
the relative phases of the eigenstates become random
and off-diagonal contributions in $\braket{\psi_t|A|\psi_t}$ vanish.

Thus, the microcanonical ensemble is applicable if $\langle A \rangle^\text{ME} = \langle A \rangle^\text{LT}$,
where the microcanonical ensemble average is defined as follows:
\begin{align}
\langle A \rangle^\text{ME} = N_\text{state}^{-1}\sum_{E_i \in [E-\delta,E+\delta]}\braket{E_i|A|E_i}.
\end{align}
Here, $E\equiv\braket{\psi_0|H|\psi_0}$ is the total energy of the system,
$\delta$ is a macroscopically small energy width,
and $N_\text{state}$ is the number of the eigenstates in the microcanonical window $[E-\delta,E+\delta]$.
For the sake of convenience, we assume that the initial state has no weight outside the microcanonical window,
\textit{i.e.},
$C_i=0$ if $E_i \not \in [E-\delta,E+\delta]$.
To find the underlying mechanism for $\langle A \rangle^\text{ME} = \langle A \rangle^\text{LT}$,
we must clarify the behavior of the eigenstate expectation value (EEV), $\braket{E_i|A|E_i}$.
The ETH states that the EEV stays constant
over a microcanonical window in the thermodynamic limit.
However, since numerical analyses can only address finite-size systems,
there always exist some fluctuations in the EEV which we call the ETH noise.
To quantify how well the ETH holds,
we consider the variance of the ETH noise
$\sigma_A^2 \equiv N_\text{state}^{-1}\sum_{E_i \in [E-\delta,E+\delta]}
\left(\braket{E_i|A|E_i}-\langle A \rangle^\text{ME}\right)^2$
and the support $s_A \equiv \max_{E_i \in [E-\delta,E+\delta]}\braket{E_i|A|E_i}
- \min_{E_i \in [E-\delta,E+\delta]}\braket{E_i|A|E_i}$.
The ETH can then be interpreted in the weak and the strong sense
as $\sigma_A \rightarrow 0$ and $s_A \to 0$
in the thermodynamic limit, respectively~\cite{Biroli10}.
The weak ETH allows an infinitesimal fraction of eigenstates,
called rare states~\cite{Biroli10},
that make EEV's deviate from their local averages.
Since the difference between $\langle A \rangle^\text{ME}$ and $\langle A \rangle^\text{LT}$ is
bounded from above by $s_A$,
the strong ETH ensures $\langle A \rangle^\text{ME} = \langle A \rangle^\text{LT}$
for any initial state.
On the other hand,
the weak ETH ensures $\langle A \rangle^\text{ME} = \langle A \rangle^\text{LT}$
for initial states satisfying $Q=O( N_\text{state}^{-1})$
because
the difference between $\langle A \rangle^\text{ME}$ and $\langle A \rangle^\text{LT}$ is
bounded through the Schwarz inequality as
\begin{align}\label{UBETH}
\left| \langle A \rangle^\text{ME} - \langle A \rangle^\text{LT} \right|
\le \sigma_A \sqrt{N_\text{state} Q -1 }.
\end{align}
Because $Q^{-1}$ represents an effective number of manybody eigenstates
in the initial state,
this assumption states that the initial state includes a large number of manybody eigenstates.
This condition is satisfied for quantum quenches in nonintegrable systems~\cite{Santos11}
and suggested to be satisfied for those from nonintegrable to integrable systems~\cite{Rigol11}.
However, this is not satisfied for quenches within integrable systems;
the microcanonical description actually breaks down for this case~\cite{Santos11,Mossel12}.
In this case,
the generalized microcanonical or Gibbs ensemble
are known to better predict
the long-time average than the microcanonical ensemble~\cite{
Rigol06,Rigol07,Cassidy11,Barthel08,Cazalilla06,Iucci09,Calabrese07,Eckstein08,Kollar08,Mossel12}.

{\it Model.---}
We analyze the finite-size scaling of the ETH noise
by invoking the Lieb-Liniger model,
which describes a one-dimensional Bose gas
with a delta-function interaction.
The Hamiltonian is given in units of $\hbar=2m=1$ as
\begin{equation}
H = \int_0^L \dd x \left[ \partial_x \Psi^\dagger(x) \partial_x \Psi(x)
+c\Psi^\dagger(x)\Psi^\dagger(x)\Psi(x)\Psi(x)\right],
\end{equation}
where $\Psi(x)$ is the bosonic field operator,
$L$ the linear dimension of the system,
$c$ the strength of the contact interaction,
and the periodic boundary condition is assumed. 
The $N$-body eigenstates are constructed from
the monodromy matrix~\cite{KorepinBook} given by
\begin{align}
&\begin{pmatrix}
	A(\lambda) & B(\lambda)\\
	C(\lambda) & D(\lambda)
 \end{pmatrix}\notag\\
&\qquad \equiv \ : \mathcal{P} \exp \left[ -\ii \int_0^L \dd x
 	\begin{pmatrix}
	\lambda /2 & \sqrt{c} \Psi^\dagger (x)\\
	-\sqrt{c} \Psi(x) & -\lambda/2
	\end{pmatrix}
	 \right] :,
\end{align}
where 
$\mathcal{P}$ and $: \cdots :$ denote path ordering and 
normal ordering of bosonic field operators, respectively.
The $N$-body eigenstate can be
obtained by operating $B(\lambda)$ $N$-times on the Fock vacuum $\ket{0}$,
$\ket{\{ \lambda_j\} } \equiv
\prod_{j=1}^N B(\lambda_j) \ket{0}$,
provided that the set of parameters 
$\{ \lambda_j \}_{j=1}^N$ satisfy the Bethe equations
\begin{align}
\ee^{\ii \lambda_j L} = - \prod_{k=1}^N
\frac{\lambda_j-\lambda_k + \ii c}{\lambda_j-\lambda_k -\ii c}
\quad (j=1,2,\dots,N).\label{Bethe_eqs}
\end{align}
Since the energy and momentum of this state are
given by
$E = \sum_{j=1}^N \lambda_j^2$
and $P = \sum_{j=1}^N \lambda_j$,
the set of parameters
$\{\lambda_j\}_{j=1}^N$ may be interpreted as representing
the momenta of the dressed noninteracting particles.

{\it Correlation Function.---}
We consider the real and imaginary parts of the EEV of the quantity
$\Psi^\dagger(x)\Psi(0)$,
{\it i.e.},
$\braket{ \{ \lambda_j \} | \Psi^\dagger(x)\Psi(0) | \{\lambda_j\} }$.
These correspond to the EEVs of Hermitian operators
$\{ \Psi^\dagger (x), \Psi(0)\}/2$ and $[ \Psi^\dagger (x), \Psi(0)]/2\ii$, respectively.
They have also been measured experimentally~\cite{Bloch00}.
These quantities are of great importance because they reflect the off-diagonal long-range order.
Substituting $\Psi^\dagger(x)=\ee^{-\ii Px}\Psi(0)\ee^{\ii Px}$
and inserting the completeness relation for the $(N-1)$-body eigenstates,
we obtain
\begin{align}\label{cor_fn}
&\braket{ \{ \lambda_j \} | \Psi^\dagger(x)\Psi(0) | \{\lambda_j\} }\notag\\
&=\sum_{\{\mu_k\}_{N-1}}\ee^{-\ii(\sum_j \lambda_j-\sum_k\mu_k)x}
\left|\braket{ \{\mu_k\} | \Psi(0) | \{\lambda_j\} }\right|^2,
\end{align}
where the summation is taken over all the $(N-1)$-body eigenstates.
The form factor $\braket{ \{\mu_k\} | \Psi(0) | \{\lambda_j\} }$
is reduced to $N$ scalar products
$\braket{ \{\mu_k\} | \{\lambda_j\}_{j\neq l} }$ $(l=1,2,\dots, N)$~\cite{Izergin87},
which are represented by the determinants of $(N-1)\times(N-1)$ matrices~\cite{Slavnov89}.
The crucial observation is that
these determinants can be summed up,
giving a single determinant of an $N\times N$ matrix~\cite{Kojima97}.
This algebraic manipulation greatly reduces the computational task,
enabling us to conduct a finite-size scaling
for $N$ as large as 35.

\begin{figure}
\begin{center}
\includegraphics[width=7cm]{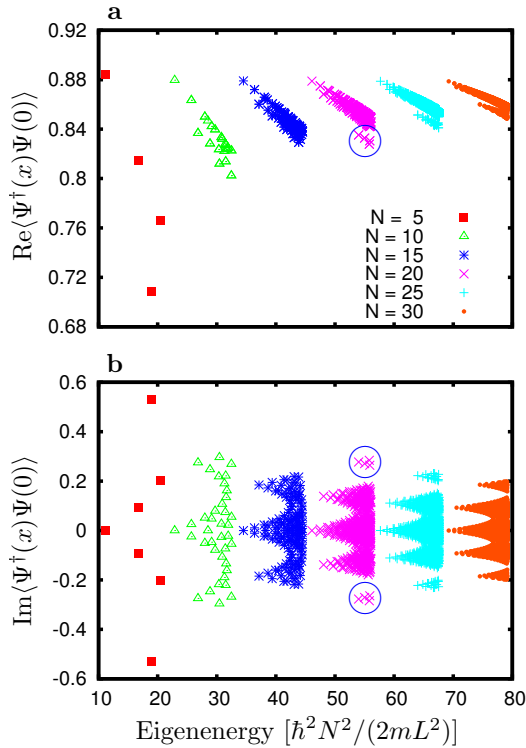}
\caption{(Color online)
(a) Real and (b) imaginary parts
of the expectation values
of $\Psi^\dagger (x)\Psi(0)$
plotted against the eigenenergy
for $N=5$ (square), $10$ (triangle), $15$ (asterisk),
$N=20$ (cross), $N=25$ (plus), and $N=30$ (dot),
where $x=L/2N$.
For each $N$,
all the energy eigenstates
in the energy window $[E_\text{g},E_\text{g}+10]$ are shown,
where $E_\text{g}$ is the ground-state energy.
The fluctuations of the data around their mean values,
which we call the ETH noise,
become smaller
with increasing the size of the system.
However, a pair of side branches enclosed by circles emerge for $N=20$
which causes the jump in Fig. 2 (see the text for details).
}
\label{real_imaginary}
\end{center}
\end{figure}

By increasing $N$ and $L$ with their ratio held fixed,
we calculate the real and imaginary parts of the quantities
$\braket{ \{ \lambda_j \} | \Psi^\dagger(x)\Psi(0) | \{\lambda_j\} }$,
as plotted in Fig.~\ref{real_imaginary}.
We set the parameters as $c=10$ and $x=1/2$,
where the unit of length is taken to be
the mean distance between particles $L/N$.
The energy window is taken as $[E_\text{g},E_\text{g}+10]$,
where, \textit{i.e.}, $E_\text{g}$ is the ground-state energy
which depends on the size of the system~\cite{window}.
The summation on the RHS of Eq.~\eqref{cor_fn} is taken over
the $(N-1)$-body eigenstates with energy up to 25.
Then, all the data have been obtained with accuracy over 97\%.
We have plotted all energy eigenstates in the energy window.
The analyses of the ETH based on these data go as follows.

\begin{figure}
\begin{center}
\includegraphics[width=7cm]{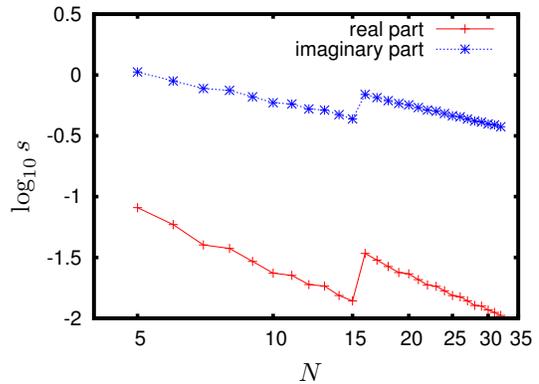}
\caption{(Color online)
Support of the ETH noise for real (plus) and imaginary (asterisk) parts
of the correlation function as functions of the size of the system
between $N=5$ and $N=32$.
The jumps at $N=16$ originates from rare states
which prevent the supports from decreasing monotonically
and invalidates the strong ETH. See the text for details.
}
\label{supp}
\end{center}
\end{figure}

First, we consider the support of the ETH noise to examine the strong ETH.
We fit each data with the least-squares method
and subtract the fitted line from the data.
Then, we define the support of the ETH noise, $s$, by
the difference between the maximum and minimum of the subtracted data~\cite{comm_window}.
The support is plotted against the number of particles in Fig.~\ref{supp},
The data shows that,
for both of the real and imaginary parts,
there exist jumps between $N=15$ and $N=16$,
which prevent the supports from decreasing monotonically.
The origin of the jumps can be seen in the EEV's for $N=20$
shown in Fig.~\ref{real_imaginary},
where a pair of side branches indicated by circles emerge.
Such branches begin to appear at $N=16$ which explains a sudden jump in Fig.~\ref{supp}.
We may expect similar side branches to appear as we increase the system size.
These additional branches, which are constituted from what are called rare states~\cite{Biroli10},
invalidate the strong ETH.

\begin{figure}
\begin{center}
\includegraphics[width=7cm]{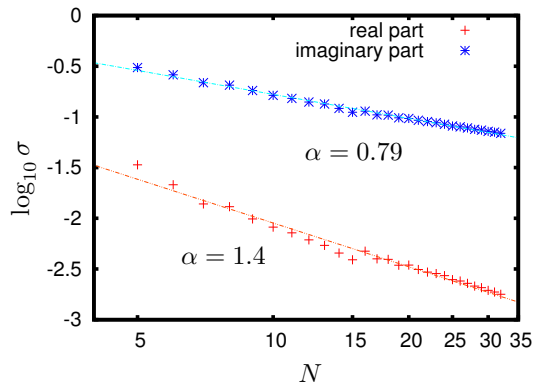}
\caption{(Color online)
Variance of the ETH noise for real (plus) and imaginary (asterisk) parts
of the correlation function as functions of the size of the system
between $N=5$ and $N=32$.
The ETH noise decays as $\sigma \propto N^{-\alpha}$,
where the least-squares fit gives
$\alpha=1.43\pm 0.04$ and $0.786\pm 0.008$ for the real and imaginary parts, respectively.
}
\label{ethnoise}
\end{center}
\end{figure}

Second, we consider the variance of the data to examine the weak ETH.
Since both the real and imaginary parts distribute around lines,
we identify the variance of the data $\sigma$
as the residual error of the least-squares method~\cite{comm_window}.
The dependence of $\sigma$ on the size of the system
is illustrated in Fig.~\ref{ethnoise}, which
shows that
the ETH noise decays with the number of the particles as $\sigma \propto N^{-\alpha}$,
where the exponent $\alpha$ is determined by the least-squares fit to be
$\alpha=1.43\pm 0.04$
and $0.786\pm 0.008$
for the real and imaginary parts, respectively.
These results show that the weak ETH
becomes better satisfied as we approach the thermodynamic limit,
despite the fact that the Lieb-Liniger model is integrable,
whereas previous studies~\cite{Rigol08,Rigol09,Rigol10,Santos10}
showed that the ETH holds worse in integrable systems than
in non-integrable chaotic systems at a fixed size of the system.
Our result implies that, if the inverse participation ratio of the initial state
is sufficiently large,
the microcanonical ensemble is applicable also to integrable systems
in the thermodynamic limit.
It is known that
we can construct a special class of initial states
which give near-thermal distributions
in integrable systems
but conserved quantities do not become arbitrarily close to
thermal expectation values even in the thermodynamic limit.~\cite{Rigol11,Chung12}.
On the other hand,
our results show that
expectation values of non-conserved quantities converge to
the thermal expectation values as we increase the system size.


{\it Interplay between ETH and Typicality.---}
Now that we have identified the power-law scaling of the ETH noise,
we can discuss which of the weak ETH and
the typicality~\cite{Popescu06,Goldstein06,Sugita06,Reimann07}
dominantly contributes to thermalization.
The typicality concerns the universal statistical-mechanical properties that
are shared by randomly generated quantum states.
Here,
we follow the typicality argument
and show that
the long-time average is close to the microcanonical ensemble average
for almost all initial states.
We assume that
the initial state is a linear combination of the manybody eigenstates
in an energy window:
$\ket{\psi}=\sum_{i=1}^{N_\mathrm{state}} C_i \ket{E_i}$,
where the set of the coefficients
$\{C_i\}_{i=1}^{N_\text{state}}$ satisfy the normalization condition
$\sum_{i=1}^{N_\mathrm{state}} |C_i|^2=1$.
Namely,
$2N_\mathrm{state}$ parameters
$\mathrm{Re}\, C_i$ and $\mathrm{Im}\, C_i$ $(i=1,2,\dots,N_\mathrm{state})$
can be represented as a point on the 
$(2N_\mathrm{state}-1)$-dimensional sphere whose radius is 1.
We denote the average over the uniform measure on the high-dimensional sphere
as an overline.
For example,
$\overline{C_i}=0$ and $\overline{|C_i|^2}=1/N_\mathrm{state}$.
Then,
the initial-state average of a long-time-averaged quantity is given by
$\overline{\langle A\rangle^{\mathrm{LT} } }=\langle A\rangle^{\text{ME}}$,
and the variance is obtained as
\begin{align}
V_A = \overline{ [ \langle A\rangle^{\mathrm{LT}}-\langle A\rangle^{\mathrm{ME} }]^2}
=\frac{\sigma_A^2}{N_\text{state}+1},
\end{align}
where $\sigma_A^2 \equiv N_\text{state}^{-1}\sum_{E_i \in [E-\delta,E+\delta]}
\left(\braket{E_i|A|E_i}-\langle A \rangle^\text{ME}\right)^2$.
Thus,
the typical magnitude of the difference between the long-time average
and the microcanonical ensemble average is given by 
$\sqrt{V_A}\sim \sigma_A N_\text{state}^{-1/2}$.
Since
$\sigma_A \propto N^{-\alpha}$ and $N_\text{state}\sim \mathrm{e}^N$,
the scaling of the typical magnitude of the difference turns out to be
\begin{align}\label{log_correction2}
\exp \left( -\frac{1}{2}N - \alpha \ln N \right).
\end{align}
The first and second terms in the exponent
describe the contributions from the typicality and the ETH,
respectively.
In this sense,
the ETH contributes to thermalization
at most as a logarithmic correction to the typicality.

{\it Conclusions and Discussions.---}
In this Letter, we have conducted a finite-size scaling analysis of the ETH
by applying the algebraic Bethe ansatz method to the Lieb-Liniger model,
and shown that the weak ETH holds in the sense that
the variance of the ETH noise vanishes as a power law in the number of particles of the system.
The weak ETH
does not necessarily mean the applicability of the microcanonical ensemble
for any initial state
due to rare states~\cite{Biroli10}.
We have shown that the microcanonical ensemble is applicable even in integrable systems
if the inverse participation ratio of the initial state is sufficiently large.
The thermalization problem for the case in which the inverse participation ratio
is not that large remains an open question.

The finite-size scaling analysis has enabled us to study the quantitative relations
between the ETH and yet another scenario for thermalization, the typicality.
We find that
the contribution of the ETH to thermalization
is at most a logarithmic correction to that of the typicality.
This originates from the fact that
the ETH is the effect of the order of some powers of
the degree of freedom of the system,
whereas the typicality argument utilizes the immense dimensionality
of the Hilbert space that grows exponentially with increasing the
degrees of freedom.
The scaling relation~\eqref{log_correction2}
which includes the ETH as a small correction to other factors 
is obtained by invoking the properties of
energy eigenstates.
However,
the fact that
it has been derived by the typicality (see Eq.~\eqref{log_correction2}), which relies
only on the immense dimensionality of the Hilbert space
with a uniform Haar measure,
strongly suggests that
the obtained relation holds quite generally.
This finding merits further study.

{\it Acknowledgements.---}
This work was supported by
KAKENHI 22340114, 
a Grant-in-Aid for Scientific Research on Innovation Areas
``Topological Quantum Phenomena'' (KAKENHI 22103005),
a Global COE Program ``the Physical Sciences Frontier'',
and the Photon Frontier Network Program,
from MEXT of Japan.
T. I. acknowledges the JSPS for financial support (Grant No. 248408).


\begin{thebibliography}{99}
\bibitem{Polkovnikov11}
	A. Polkovnikov, K. Sengupta, A. Silva, and M. Vengalattore,
	Rev. Mod. Phys.
	\textbf{83}, 863 (2011).
\bibitem{Yukalov12}
	V. I. Yukalov,
	Laser Phys. Lett.
	\textbf{8}, 485 (2011).
\bibitem{Kinoshita06}
	T. Kinoshita, T. Wenger, and D. S. Weiss,
	Nature 440, 900-903 (2006).
\bibitem{Trotzky11}
	S. Trotzky, Y. -A. Chen, A. Flesch, I. P. McCulloch,
	U. Schollw\"{o}ck, J. Eisert, and I. Bloch,
	Nature Physics \textbf{8}, 325 (2012).
\bibitem{Tasaki98}
	H. Tasaki,
	Phys. Rev. Lett.
	\textbf{80}, 1373 (1998).
\bibitem{Goldstein06}
	S. Goldstein, J. L. Lebowitz, R. Tumulka, and N. Zanghi,
	Phys. Rev. Lett.
	\textbf{96}, 050403 (2006).
\bibitem{Reimann08}
	P. Reimann,
	Phys. Rev. Lett.	
	\textbf{101}, 190403 (2008).
\bibitem{Rigol08}
	M. Rigol, V. Dunjko, and M. Olshanii,
	Nature (London)
	\textbf{452}, 854 (2008).
\bibitem{Gogolin11}
	C. Gogolin, M. P. M\"{u}ller, and J. Eisert,
	Phys. Rev. Lett.
	\textbf{106}, 040401 (2011).
\bibitem{Ikeda11}
	T. N. Ikeda, Y. Watanabe, and M. Ueda,
	Phys. Rev. E
	\textbf{84}, 021130 (2011).
\bibitem{Popescu06}
	S. Popescu, A. J. Short, and A. Winter,
	Nature Physics
	\textbf{2}, 754 (2006).
\bibitem{Sugita06}
	A. Sugita,
	RIMS Kokyuroku (in Japanese)
	\textbf{1507}, 147 (2006).
\bibitem{Reimann07}
	P. Reimann,
	Phys. Rev. Lett.
	\textbf{99}, 160404 (2007).
\bibitem{Short11}
	A. J. Short,
	New Journal of Physics
	\textbf{13}, 053009 (2011).
\bibitem{Deutsch91}
	J. M. Deutsch,
	Phys. Rev. A
	\textbf{43}, 2046 (1991).
\bibitem{Srednicki94}
	M. Srednicki,
	Phys. Rev. E
	\textbf{50}, 888 (1994).
\bibitem{Horoi95}
	M. Horoi, V. Zelevinsky, and B. A. Brown,
	Phys. Rev. Lett.
	\textbf{74}, 5194 (1995).
\bibitem{Rigol09}
	M. Rigol,
	Phys. Rev. Lett.
	\textbf{103}, 100403 (2009).
\bibitem{Rigol10}
	M. Rigol and L. F. Santos,
	Phys. Rev. A
	\textbf{82}, 011604(R) (2010).
\bibitem{Santos10}
	L. F. Santos and M. Rigol,
	Phys. Rev. E
	\textbf{82}, 031130 (2010).
\bibitem{Lieb63}
	E. H. Lieb and W. Liniger,
	Phys. Rev.
	\textbf{130}, 1605-1616 (1963);
	E. H. Lieb,
	Phys. Rev.
	\textbf{130}, 1616-1624 (1963).
\bibitem{Biroli10}
	G. Biroli, C. Kollath, and A. M. L\"{a}uchli,
	Phys. Rev. Lett.
	\textbf{105}, 250401 (2010).
\bibitem{Rigol06}
	M. Rigol, A. Muramatsu, and M. Olshanii,
	Phys. Rev. A
	\textbf{74}, 053616 (2006).
\bibitem{Rigol07}
	M. Rigol, V. Dunjko, V. Yurovsky, and M. Olshanii,
	Phys. Rev. Lett.
	\textbf{98}, 050405 (2007).
\bibitem{Cassidy11}
	A. C. Cassidy, C. W. Clark, and M. Rigol,
	Phys. Rev. Lett.
	\textbf{106}, 140405 (2011).
\bibitem{Barthel08}
	T. Barthel and U. Schollw\"{o}ck,
	Phys. Rev. Lett.
	\textbf{100}, 100601 (2008).
\bibitem{Cazalilla06}
	M. A. Cazalilla,
	Phys. Rev. Lett.
	\textbf{97}, 156403 (2006).
\bibitem{Iucci09}
	A. Iucci and M. A. Cazalilla,
	Phys. Rev. A 
	\textbf{80}, 063619 (2009).
\bibitem{Calabrese07}
	P. Calabrese and J. Cardy,
	J. Stat. Mech.,
	P06008 (2007).
\bibitem{Eckstein08}
	M. Eckstein and M. Kollar,
	Phys. Rev. Lett.
	\textbf{100}, 120404 (2008).
\bibitem{Kollar08}
	M. Kollar and M. Eckstein,
	Phys. Rev. A
	\textbf{78}, 013626 (2008).
\bibitem{Mossel12}
	J. Mossel and J. -S. Caux,
	New J. Phys.
	\textbf{14} 075006 (2012).
\bibitem{Santos11}
	L. F. Santos, A. Polkovnikov, and M. Rigol,
	Phys. Rev. Lett.
	\textbf{107}, 040601 (2011).
\bibitem{Rigol11}
	M. Rigol and M. Srednicki,
	Phys. Rev. Lett.
	\textbf{108}, 110601 (2012).
\bibitem{KorepinBook}
	V. E. Korepin, N. M. Bogoliubov, and A. G. Izergin,
	\textit{Quantum inverse scattering method and
	correlation functions},
	Cambridge (1997).
\bibitem{Bloch00}
	I. Bloch, T. W. H\"{a}nsch, and T. Esslinger,
	Nature
	\textbf{403}, 166 (2000).
\bibitem{Izergin87}
	A. G. Izergin, V. E. Korepin, and N. Y. Reshetikhin,
	J. Phys. A
	\textbf{20}, 4799-4822 (1987).
\bibitem{Slavnov89}
	N. A. Slavnov,
	Theor. Math. Phys.
	\textbf{79}, 502-508 (1989).
\bibitem{Kojima97}
	T. Kojima, V. E. Korepin, and N. A. Slavnov,
	Commun. Math. Phys.
	\textbf{188}, 657-689 (1997).
\bibitem{window}
	We set the microcanonical window not by energy per particle
	but by energy itself for the following two reasons.
	First, the energy window is closely related to
	the finiteness of the resolution of experimental apparatuses to measure energy
	which is evaluated by the absolute value of energy.
	Second, it is not taken as energy per particle in most of the references,
	for example Ref.~\cite{Rigol08}.
	However, the essential results are not changed whichever notation is used.
\bibitem{comm_window}
	The microcanonical window should, in principle,
	be chosen so small that the microcanonical average is almost independent of the width,
	but our choice, $[E_\text{g}, E_\text{g} + 10]$, does not satisfy this condition.
	The width is chosen for a mathematical trick
	which enables us to utilize more data points.
	After subtracting the systematic linear behaviors as we do in the manuscript,
	we can address a decent number of data points to obtain the variance of the EEV's.
	This trick does not essentially change our results.
\bibitem{Rigol11-2}
	M. Rigol and M. Fitzpatrick,
	Phys. Rev. A
	\textbf{84}, 033640 (2011).
\bibitem{Chung12}
	M. -C. Chung, A. Iucci and M. A. Cazalilla,
	New Journal of Physics
	\textbf{14} 075013 (2012).
\end{thebibliography}
\end{document}